\documentclass[aps,prl,twocolumn,superscriptaddress,groupedaddress]{revtex4}
\usepackage{subfig}
\usepackage{graphicx}  % needed for figures
\usepackage{dcolumn}   % needed for some tables
\usepackage{bm}        % for math
\usepackage{amssymb}   % for math
\usepackage{slashed}
\usepackage{graphicx}				% Use pdf, png, jpg, or eps with 	
\usepackage{amsmath}
\usepackage{mathtools}
\usepackage{tikz,pgf}
\usepackage{comment}
\usepackage{asymptote}
\usetikzlibrary{arrows,backgrounds}
\usetikzlibrary{fit,scopes,calc,matrix,positioning,decorations.pathmorphing}
\usepackage[all]{xy}
\usepackage{yfonts}
\newcommand{\bra}[1]{\ensuremath{\left\langle#1\right|}}
\newcommand{\ket}[1]{\ensuremath{\left|#1\right\rangle}}

\begin{document}
\title{Gauge is quantum?}
\author{Andrei T. Patrascu}

\begin{abstract}
In this article I discuss a proposed equivalence between quantum mechanics and gauge theory. The evolution of the idea of gauge invariance is very interesting to follow even historically. The idea that gauge transformations have in general no impact on the physics described, leading to even calling gauge variant observables as "unphysical" has transformed quite significantly in the past. Now we know large gauge transformations become sensitive to topological (homotopical) structures of the underlying group, as is the case of the SU(2) anomaly, and that gauge degrees of freedom do play a role, for example in the problem of confinement, via the Gribov ambiguity. The evolution of the concept of gauge is continued here, by the main claim, that gauge and quantum have equivalent origins and behave similarly to the point that they can be considered dual representations of the same physical ideas. 
\end{abstract}
\maketitle
\section{Introduction}
An interesting phenomenon is happening in the construction of the Madelung equations from the Schrodinger equation. It seems like the Madelung equations require a rotational invariance symmetry to properly account for quantum vortices, and that Madelung equations are not fully determining the dynamics. The relation between Schrodinger's equations and Madelung equations is often debated with the observation that no clear understanding exists for why the additional rotational discretisation condition is required. Here I explain it as an additional gauge symmetry that speaks in favour of the recent idea that quantum is gauge (Q=G). Indeed, this additional symmetry seems to emerge as a gauge symmetry condition that needs to be incorporated in the Madelung equations in order to properly describe quantum mechanics. In that sense "Madelung Equation + Gauge symmetry = Quantum mechanics". Arguments in favour of understanding the quantum phase as a gauge symmetry component of the solution of Schrodinger's equations are also introduced. It is interesting to note that non-local observables emerge in all theories of gauge interactions due to some curvature. In general relativity, all observables that are gauge invariant must be non-local hence we cannot fully localise general relativistic observables. In gauge theories like Quantum Chromodynamics (QCD) in order for observables to be gauge invariant they also need to be non-local, only this time in the gauge space, allowing QCD to be fully localised in spacetime while forcing us to consider the colour degrees of freedom as gauge variant unless non-locally defined in that gauge space (forcing us to consider only "white" states as unambiguous). Quantum mechanics also has its own form of non-locality, which results from a global construction of the wavefunction. The relative phases of the wavefunction components can probe the non-local structure of the system, but they only become detectable if a probabilistic approach is implemented and we measure a statistics of the experimental outcomes. This effect must also correspond to some form of curvature. Given that I do not support the ideas that quantum mechanics is emergent or that there exists a duality between classical and quantum physics, I rise the question as of what such a curvature could be and in what space could it be observed? As an additional curvature term emerges, it becomes clear that quantum behaviour becomes relevant as it depends on two scales, one is the well known scale defined by the Planck constant, the other by the quantum curvature. In the same way in which we expect to approximate local gravitational observables when the spacetime curvature is low, and we can expect to see cvasi-free quarks when the QCD curvature is low (at high energies) in quantum mechanics we may have an interplay between $\hbar$ and the curvature, having quantum behaviour appear even when $\hbar$ is relatively small, given a large enough quantum curvature, or having classical behaviour even when $\hbar$ is relevant, given a small curvature.
%%%

Historically, quantum mechanics appeared because of the realisation that a certain experimental setup does not fully determine all observables that we can imagine about a system. Before quantum mechanics, it was assumed that there are universal properties that can be fully determined about any system, independent of the construction of the system, the context it is in, or the experimental setup. This assumption was wrong. The simplest realisation was that an experimental setup incorporating a physical system in which position is fully determined does not allow for a perfect determination of the momentum. This is why momentum had to be represented as a matrix observable with a series of eigenvalues representing the possible outcomes of the momentum given that the experiment itself couldn't determine them. 
\begin{figure}
  \includegraphics[width=\linewidth]{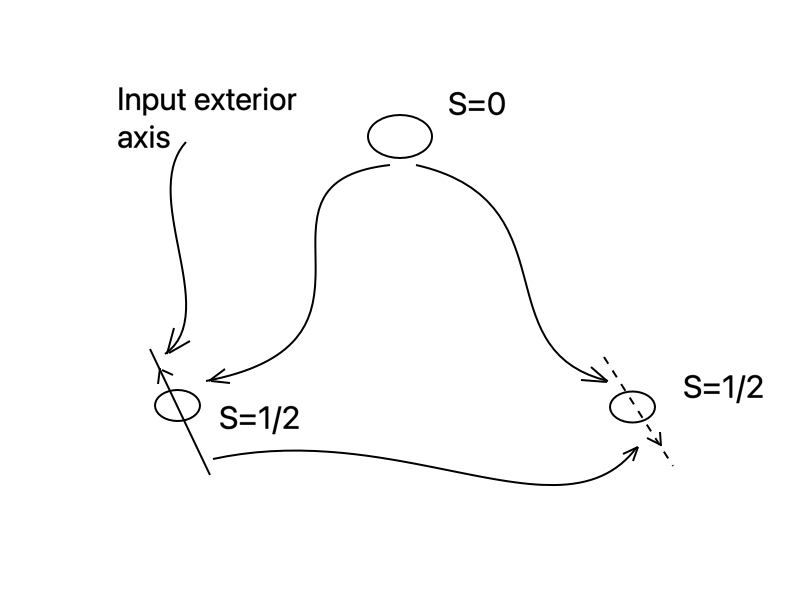}
  \caption{The spin 0 particle decays into two spin 1/2 particles. Without a predefined axis on the two decay products, nothing can be said about a spin projection. There exists a full symmetry linking all possible spin $\pm 1/2$ orientations. Once parallel axes are introduced from outside, a frame is chosen, and correlations between the two outcomes can be observed, while the general initial symmetry between all possible axis orientations is eliminated. }
  \label{fig:gauge1}
\end{figure}
Another example, maybe more suitable for the context of this article is a spin $0$ system that decays into two electrons of spin $1/2$ (see fig.1). Various conservation rules make us know that one of the electrons will have spin $1/2$ and the other would have the opposite spin. However, there are some missing things in the presentation above. First, we need to talk about spin projection, not spin itself, which remains $1/2$. Second, in order to discuss about spin projection we need to discuss about a spin projection axis, which was neither introduced in the experimental setup nor mentioned in the initial definition of the spin $0$ system. Nature doesn't provide such an axis, hence it remains undetermined. That means that in order to measure and test the claim that the "spins will be opposite" we need to introduce an arbitrary axis, on both electron positions, and measure the spin projection on them. If we introduce the two axes parallel to each other, we will indeed obtain opposite spin projections. But we may have chosen the $z$ axis first, and obtain opposite spin projections $1/2$, and then we may choose the $x$ axis and obtain again opposite spin projections of the magnitude $1/2$. That shows that the system had no actual realisation of the spin projection to begin with. This property was arbitrary unless we introduce an axis. This means that the observables we defined initially could not have determined the properties of the spin projection uniquely, and that is to be expected, as no axis for the spin projection was given or inherent to the system initially. The evolution of the system had to propagate this arbitrary function of a choice of an axis across time. This "redundant" information about oppositeness of the spin projections no matter what axis is chosen is encoded in the complex phase of the solution of the Schrodinger equation. The same type of arbitrariness is also included in another type of systems, namely in gauge systems. The claim here is that gauge and quantum systems have a common origin and that in fact they are fundamentally the same. With this result, one can come up with, for example, quantum entanglement relating phase-containing wave-functions, that is expected to have some dual or analogue in the gauge dynamics of gauge systems, and indeed, it does. That will lead to new dualities between quantum information aspects and gauge systems. But let me now describe the idea behind gauge systems properly. It seems that most physical systems allow for more degrees of freedom than the dynamical ones, and while they often seemed spurious, nobody was able to get rid of them when studying for example interacting systems. While it is possible to eliminate all gauge degrees of freedom, the resulting theory looses predictive power unless we know in advance what the role of the interactions would be. In fact, gauge degrees of freedom prove essential to the description of a series of problems, for example Chern-Simons theory, etc. aside of the usual theories of interacting quantum fields [1,2,3]. In any case, what happens is that in general we expect the equations of motion to provide us with a unique solution for the dynamics. However, this only happens provided a certain function to be discussed later on, is invertible. This does not happen in general, leading to a multivariate problem in which arbitrary functions emerge in the solutions of the dynamical equations. Those arbitrary functions represent the arbitrariness of the choice of a gauge. A gauge is like a frame, be it a reference frame in spacetime or a frame in some inner space of the theory. In any case, such a choice must remain arbitrary if we are to obtain consistent solutions. But the fact that there is no absolute reference frame and the choice of a reference frame is arbitrary doesn't make reference-frame variant observables any less physical. After all three dimensional length is not a relativistic invariant but it remains an observable physical property for a respective reference frame. I will discuss about the gauge equivalent of this later on. In any case, arbitrary functions appear in the solutions of dynamical systems.With them, we allow for various possible selections of frames at any arbitrary instance of time. This characteristic is of course not directly observable, but if we ignore gauge freedom we would have a very hard time describing interaction and propagation of interactions. In fact the same situation occurs in quantum mechanics. We know that the phase of a quantum wavefunction is not directly observable
\begin{equation}
\Psi(x,t)=e^{i\hbar \phi}\mathcal{A}(x,t)
\end{equation}
results in 
\begin{equation}
|\Psi(x,t)|^2=|\mathcal{A}(x,t)|^{2}
\end{equation}
because of course 
\begin{equation}
|e^{i\hbar\phi}|^{2}=(e^{i\hbar\phi})^{*}(e^{i\hbar\phi})=e^{-i\hbar\phi}\cdot e^{i\hbar\phi}=1
\end{equation}
This makes the phase of the overall wavefunction invisible and hence arbitrary. However, that arbitrary function makes also possible, when we have a relative phase between various wavefunction components, to obtain interference that does indeed manifest itself in the shape of $\rho(x,t)=|\Psi(x,t)|^{2}$ and gives us access to non-local information. 
The use of the term "non-local" must be slightly refined to make it clearer what is meant by it. Unfortunately the terms local and causal are often used interchangeably although this would be a clear mistake. The ideas determining the meaning of these concepts are nicely analysed in the article [10] but it would make sense to repeat some of the ideas here for clarity. 
First, we should note that quantum entanglement is a fundamentally causal phenomenon. At some point in the creation of an entangled system, its parts must have interacted in some way. That interaction was causal and local. Therefore when we talk about quantum non-locality we must underline that it refers to non-local correlations that emerge in systems that do share entanglement or other types of quantum superpositions but that were in causal contact and interacted locally. From the point of view of quantum communications, we can distinguish between locality and causality in a different way, based on the following thought experiment. Let us take a physical system that is divided into two separate parts, part A controlled by an agent called Alice, and part B, controlled by an agent called Bob. Both agents share a joint quantum state whose density operator $\rho_{AB}$ is not known. The goal of the experiment is to transform this state into $\mathcal{E}(\rho_{AB})$, where $\mathcal{E}$ is a given operation. 
As the initial state is shared, the ability of Alice and Bob to construct any state by applying any operation $\mathcal{E}_{AB}$ to their state depends on the information they can transmit to each other. To study the concepts of locality and causality it is interesting to analyse what kind of operations they can implement if they cannot communicate by any means (classical or quantum). Such operations are the equivalent of the operations that can be constructed if Alice and Bob are spacelike separated. The inclusion of quantum operators, as will be seen, therefore modifies our understanding of the usual answers to the question "how can A know something about B?" or "how can we know about something happening somewhere else?". Alice and Bob will be able to use a shared entangled ancilla state that could have been prepared in advance. They both are able to perform measurements on their sides, but they cannot know each other's experimental results. Because of this it is usually common to consider the so called trace preserving superoperators where no post-selection of the quantum state based on the measurement outcome is allowed. A bipartite superoperator is localisable if it can be implemented by Alice and Bob acting only locally on the shared state and the shared ancilla but without allowing any communication between them. Causality however is defined differently. Such a superoperator is called causal if it doesn't allow any of the two agents to send a signal to each other. If for example Bob applies a local superoperator $\mathcal{B}$ to his subsystem that is part of the shared system, just before the global superoperator $\mathcal{E}$ acts, and Alice makes a local measurement on her half just after $\mathcal{E}$ has acted, then if the measurement performed by Alice can gain any information about what operation was applied by Bob, then Bob was able to signal Alice. The operation $\mathcal{E}$ is causal if no such signalling is possible in either direction. We know that entanglement cannot be used to send a superluminal signal from Alice to Bob or from Bob to Alice. This means that every localisable superoperator must be causal. However, causal superoperators may not be always localisable. 
A superoperator can be interpreted as a generalised measurement with an unknown outcome, acting on a density operator $\rho$ and having a representation of the form 
\begin{equation}
\mathcal{E}(\rho)=\sum_{\mu}M_{\mu}\rho M_{\mu}^{\dagger}
\end{equation}
where the normalisation condition becomes 
\begin{equation}
\sum_{\mu}M_{\mu}^{\dagger}M_{\mu}=I
\end{equation}
In general $tr(\mathcal{E}(\rho))$ is interpreted as the probability of the observed outcome. 
Every superoperator has a unitary representation. If we define a Hilbert space $\mathcal{H}_{S}$, and we want to implement the superoperator $\mathcal{E}_{S}$ acting on that Hilbert space, we introduce an ancilla with Hilbert space $\mathcal{H}_{R}$, prepare the pure state $\ket{\psi}\in\mathcal{H}_{R}$ of the ancilla, and perform an unitary transformation $U$ on $\mathcal{H}_{S}\otimes \mathcal{H}_{R}$, following by the elimination of the ancilla
\begin{equation}
\mathcal{E}_{S}(\rho_{S})=tr_{R}[U(\rho_{S}\otimes \ket{\psi}_{R}\;_{R}\bra{\psi})U^{\dagger}]
\end{equation}
If the bipartite superoperator $\mathcal{E}$ does not allow superluminal signalling from $B$ to $A$ we say that the superoperator is only semi-causal. This is necessary as there exist unidirectional barriers like black hole horizons. If signalling is impossible in both directions, we call the operator causal. Let us return at our system formed out of two subsystems. In that case, Alice may control not only the subsystem A, but also the ancilla system R while Bob controls the subsystem B and the ancilla system S. Initially, Alice and bob could share an initial density operator $\rho_{RABS}$ in the Hilbert space $\mathcal{H}_{R}\otimes \mathcal{H}_{A}\otimes \mathcal{H}_{B}\otimes \mathcal{H}_{S}$ and let us consider that Bob applies a superoperator $\mathcal{B}_{BS}$ on his half of the state. After the operation $\mathcal{E}$ is performed, we obtain Alice's density operator by tracing out Bob's system and ancilla
\begin{equation}
\rho_{RA}=tr_{BS}[\mathcal{E}_{AB}((I_{RA}\otimes \mathcal{B}_{BS}(\rho_{RABS}))]
\end{equation}
If this final state of Alice depends in any way on the superoperator applied by Bob, $\mathcal{B}$ then we have non-zero mutual information between Bob's superoperator and Alice's measurement. Therefore we can have Bob transmit Alice classical information. A system is called semi-causal if this uni-directional transmission is not possible. If the superoperator $\mathcal{E}$ is not semicausal, then there exists such an initial state $\rho_{RABS}$ shared by Alice and Bob, as well as a superoperator $\mathcal{B}$ that can be applied by Bob such that 
\begin{equation}
tr_{BS}[\mathcal{E}_{AB}(\rho_{RABS}]\neq tr_{BS}[\mathcal{E}_{AB}(I_{RA}\otimes \mathcal{B}(\rho_{RABS}))]
\end{equation}
Now that we have a concept of causality, let us see the idea of localisability and how the two concepts differ. A physical system is seen as local, but as it is, it can be formed by many subsystems that may be acted upon locally and independently of each other. The operations that can applied to such a system at a fixed time without communication among local agents are localisable. Nevertheless, the agents, while not communicating, can prepare states in advance. They can for example share an ancilla that could be an entangled quantum state and consume the shared entanglement while executing their local operations. We call therefore a bipartite superoperator localisable if and only if
\begin{equation}
\mathcal{E}(\rho_{AB})=tr_{RS}[\mathcal{A}_{RA}\otimes \mathcal{B}_{BS}(\rho_{AB}\otimes \rho_{RS})]
\end{equation}
for a shared ancilla state $\rho_{RS}$ and local superoperators $\mathcal{A}_{RA}$ and $\mathcal{B}_{BS}$. 
With these definitions we understand the concepts of localisable and causal in a better way from the perspective of the quantum operations involved. This introduction to the concepts of causality, locality, and semi-causality are based on ref. [10]. Of course, the concept can be used in broader terms with the possibility that the underlying space changes and becomes our gauge space. Such modifications are a simple extension of the idea of localisability and causality in spacetime, but with different restrictions, as, for example, the gauge space has a concept of locality that can then be projected onto spacetime, while causality is defined on spacetime by definition. 
Let us now understand several aspects related to the idea of gauge variant and invariant observables as well as the meaning of the quantum phase in this context. 

The arbitrariness of the choice of a quantum phase doesn't mean the effects of the existence of a phase are not physical. The exactly same thing happens for a gauge invariance. Interestingly enough, in terms of a quantum field, it also appears as a phase factor in the fields, and the demand that any arbitrary choice of a gauge phase doesn't have an effect on the physical action results in the requirement of a covariant derivative defined according to the type of gauge symmetry introduced and in a gauge connection (or an interaction mediating gauge field). Interestingly enough, there has always been an interplay between gauge and quantum, for example in the case of quantum gauge anomalies [4,5,6,7]. In the process of Feynman path integral quantisation, the integration measure which is required to perform quantisation has in some cases an undesirable effect on the gauge symmetry of the theory itself. This shows that quantum and gauge share the same structure and therefore can affect each other. Demanding the restoration of gauge symmetry (arbitrariness of the choice of a frame) in the overall theory by various mechanisms brought several new insights into how Nature works. However, the fact that one can affect the other in such a way is the result of them being basically the same. Therefore the first claim of this paper is that indeed quantum is gauge. 
Of course, we know of gauge in several theories that we do not consider to be quantum, for example the classical Maxwell equation is definitely having a gauge invariant but it is generally considered to be classical. That may be true, but it is also possible that even a classical Maxwell equation has some aspects of quantum dynamics that were simply not identified as such when the theory was constructed, one of them being its natural gauge invariance. Another would be the physical absence of a longitudinal field component. Therefore, I consider that such theories have gauge as a form of "quantum remnant". Also, in the process of propagation of electromagnetic waves, it is impossible to identify a completely separable region of the manifold that allows for the propagation of the wave [8]. In any case, gauge transformations are transformations we could perform at the level of the action of the theory, while allowing the equations of motion to be apparently unchanged. The action however, in a Hamilton-Jacobi approach to quantisation appears in the complex phase of the resulting wavefunction, and remains with this role in all approaches to quantisation, including and in particular the Feynman path integral quantisation. We know how to make changes to the action that represent simply a gauge transformation and leave the theory invariant. The fact that quantum mechanics itself is the result of the existence of such an arbitrariness was up to now not noticed. 
As stated in the very beginning of this article, the existence of quantum mechanics is due to the fact that not all observables are fully determined in all experimental contexts. This has nothing to do with thermodynamic emergence or any thermodynamical "loss" of microscopic degrees of freedom, but instead it only depends on how the observables are being defined and therefore on how the quantum experiment is being prepared. The fact that observables do not have one single determined outcome only, in so many contexts, means that we must be prepared to carry in our solution enough arbitrary functions capable of encoding all those non-determined contexts. That is a gauge arbitrary function, which can also be seen as a quantum complex phase. 
The removal of quantum gauge anomalies just makes the two arbitrary functions introduced ultimately by the same type of gauge freedom, more compatible with each other. From this perspective the construction of a reference frame and even of a causal structure as in special relativity has a dual in quantum entanglement. 
\section{Gauge as quantum}
Before we introduce the concept of gauge in a dynamical equation, it probably is important to make some observations about what it means to be gauge invariant and gauge variant. It is often assumed that only gauge invariant observables are "physical" while any gauge variant observables are not detectable, measurable, or visible. In fact, what it means for an observable to be gauge variant is just that it depends on the gauge choice and that in principle observers making the same experimental measurement with different gauge choices will detect different results. But it also makes sense to notice that because a gauge choice was necessary to perform the experiment and obtain the result, maybe the experimental setup itself was not sufficient to determine fully the gauge "variant" observable we tried to determine, and in fact the "choice of a gauge" in a sense "completes" the experimental setup with some information that allows the system to fall onto one state or another. This choice is of course arbitrary, and in quantum mechanics there is no fundamental way of determining which of the possible observational eigenvalues will emerge. We will have to change the experimental setup to make the outcomes of one special observable strictly determined, but then we will determine them in a different context. This remains valid in gauge theory as well. The absence of a gauge choice makes the exact determination of the outcomes impossible, the addition of a gauge makes the observable able to give us results, but the actual results will remain unpredictable. The solution around this unpredictability in gauge theory is found by the introduction of additional variables that make the distinction between the possible outcomes manifest at least mathematically, however, physically the distinction is fictitious, and the variables we introduce as auxiliary are essentially mute. Mathematically we can now distinguish them, but physically they remain equivalent and will appear in the experiment only probabilistically. Therefore we may add variables to make the system mathematically one-to-one, but the target states will remain physically indiscernible. Now, those gauge variant observables encode precisely this: that we can determine them, but only two observers making the same choice of gauge will agree upon the outcomes. The outcomes remain therefore ambiguous, but whether that is an issue is more of a matter of taste than of physics. In quantum mechanics something similar happens but there we have been forced to accept it as such. The observable itself doesn't determine fully the outcome. If we add auxiliary variables we can make the outcomes of the observable mathematically distinguishable but physically they will remain undistinguishable and hence will occur each with its own probability as dictated by quantum mechanics. If however we wish to determine them perfectly, in gauge theory we need to set up a series of additional constraints. These constraints however change the dynamical system and hence the experimental setup. We will then be able to determine accurately the observable we wish but the system won't be the same. This is also why upon the measurement of a position, the momentum becomes undetermined. The constraints that require the perfect determination of the position will change the system into one in which the momentum will be fully undetermined. The same type of constraints also work in gauge theory, and the outcome is the same. We can turn the system around and make the gauge variant observable gauge invariant by various forms of dressing, but that implies the system won't be the one we intended to study to begin with. It will be a system that fully determines our previously gauge "variant" observable in an unambiguous way, but will render other observables, previously gauge invariant, now variant. 
In any case something that is gauge variant doesn't necessary have to be unphysical. This is usually assumed and postulated by many authors and became almost a universal belief of the scientific community. However, there are degrees of freedom that are gauge variant but that are totally physical in the sense that they do have an impact on the dynamics of their systems even though we cannot detect them freely. One example is the colour of the quark. An arbitrary choice of an SU(3) transformation which is a gauge symmetry of QCD transforms the colour of a quark into another colour or into a superposition of colours. In any case, quark colours are not gauge invariant, but we cannot possibly assume that they are not physical or that the theory of chromodynamics is based on something that "does not exist". Therefore gauge dependent observables and degrees of freedom are not "unphysical", they are simply ambiguous. Nature doesn't seem to have a problem with this type of ambiguity. We claim that we don't see the colour of a quark freely in nature, but then this could change in the asymptotically free limit of QCD. But this is not even necessary. What the theory says is that we could in principle see the colour of the quarks in a hadron for example, with the observation that their colour is not fully determined unless we fix an arbitrary gauge. Designing experiments that "conspire" to have the same choice of a gauge would reveal the same colour. Colour remains an ambiguous degree of freedom, but certainly not an "unphysical" one. 
In general when we describe the evolution of a classical dynamical system we start with an action functional 
\begin{equation}
S_{L}=\int_{t_{1}}^{t_{2}}L(q,\dot{q})dt
\end{equation}
and obtain the stationary solutions of this action under variations $\delta q^{n}(t)$ of the Lagrangian variables that vanish at the endpoints. The condition for this stationarity appears to be given by the Euler Lagrange equations which are in their simplest form given by 
\begin{equation}
\frac{d}{dt}(\frac{\partial L}{\partial \dot{q}})-\frac{\partial L}{\partial q^{n}}=0
\end{equation}
the general expectation is that such an equation will fully determine the trajectory of the system in the space $(q,\dot{q})$ and hence the system will evolve according to a fully determined dynamics. Otherwise stated, it was expected that the evolution will be following a single well determined path or trajectory from the initial to the final state. We can re-write this equation as 
\begin{equation}
\ddot{q}^{n'}\frac{\partial ^{2} L}{\partial \dot{q}^{n'}}{\partial \dot{q}^{n}}=\frac{\partial L}{\partial q^{n}}-\dot{q}^{n'}\frac{\partial ^{2}L}{\partial q^{n'}\partial \dot{q}^{n}}
\end{equation}
In this equation we expect that the acceleration at a given time to be uniquely determined by the positions and velocities at that time. This conclusion however is subject to a condition, namely that 
\begin{equation}
det(\frac{\partial^{2}L}{\partial \dot{q}^{n}\partial \dot{q}^{n'}})\neq 0
\end{equation}
If however the determinant is indeed zero the acceleration will not be uniquely determined by the positions and velocities and that induces arbitrary functions of time in the solutions of the equations of motion. This is exactly the condition for gauge degrees of freedom to exist. To have a gauge theory we need therefore that 
\begin{equation}
\frac{\partial^{2}L}{\partial \dot{q^{n'}}\partial\dot{q}^{n}}
\end{equation}
cannot be inverted. 
Now, if we think at quantum mechanics and at the path integral formulation by Feynman, we realise that what we do is to integrate over all possible paths given by the action $S[q,p,t]$ resulting, due to the quantum nature of the trajectories, into the integration over all possible continuous, yet not continuously differentiable paths. In Feynman's path integral formulation we also cannot single out one trajectory when we compute the phase structure of our wavefunction, and therefore we calculate the integral over all possible paths. The same thing happens here, namely the function mentioned above is non-invertible, resulting in the impossibility of singling out one path as the "classical" or "true" path of the dynamical system. The non-invertibility condition of a function usually means that the function creates a map that is not one to one, namely say, a position and a velocity at a certain time will not produce one acceleration alone, or reversely. But I hope it is clear how this is exactly what happens when we decide to integrate over all possible paths, we cannot unambiguously single out one path among all possible paths, each with a different value for $S[q,p,t]$. 
Now, we can do the same thing in Hamiltonian dynamics with no loss of generality. There we start with the canonical momenta, namely 
\begin{equation}
p_{n}=\frac{\partial L}{\partial \dot{q}^{n}}
\end{equation}
the vanishing of the determinant above is just a condition for the non-invertibility of the velocities as functions of the coordinates and momenta. Therefore the momenta are not all independent in this case but are determined by some relations
\begin{equation}
\begin{array}{cc}
\phi_{m}(q,p)=0,& m=1,...,M
\end{array}
\end{equation}
this leads to an interesting observation that can be particularised again to the path integral formulation. There I mentioned that the trajectories are continuous but not continuously differentiable, which results in jumps of the variables or their conjugate partners, which is something that is inherent to an extreme case of the above condition. We see therefore how, allowing for an arbitrary function in the solution of the dynamics equations makes the system be both gauge and closer to being quantum. The common origin of the two appears more obvious. It is not the equation of motion that introduced this constraint, but the simple fact that the function $v=v(q,p)$ is non-invertible. Also, those constraints imply no restrictions on the coordinates $q$ and their velocities $\dot{q}$. We assume that the matrix $\frac{\partial^{2}L}{\partial \dot{q}^{n} \partial \dot{q}^{n'}}$ has a constant rank over the $(q, \dot{q})$ space and the constraint equations define a submanifold that is smoothly embedded in the phase space. This submanifold is the primary constraint surface in dynamical gauge theories. If we consider the rank of $\frac{\partial^{2}L}{\partial \dot{q}^{n} \partial \dot{q}^{n'}}$ to be $N-M'$ then we have $M'$ independent equations among the constraints and then the primary constraint surface forms a phase space submanifold of dimension $2N-M'$. This means that the inverse transform from $p$ to $\dot{q}$ is multivalued. For a point in phase space $(q^{n}, p_{n})$ which obeys the constraints the inverse image $(q^{n},\dot{q}^{n})$ that solves $p_{n}=\frac{\partial L}{\partial \dot{q}^{n}}$ cannot be unique since this equation of motion defines a map from a $2N$ dimensional manifold $(q, \dot{q})$ to a manifold of dimension $2N-M'$. This means that an inverse map of a given point of the submanifold $\phi_{m}(q,p)=0$ produce a manifold of dimension $M'$. In order to make the transformation single valued, one has to introduce extra parameters, at least a number $M'$ of them. These may appear as Lagrange multipliers. 
To pass to the Hamiltonian formalism some restrictions on the constraints appear. Mainly those conditions imply that the constraint surface must be coverable by open regions on each of which locally, the constraint functions can be split into the independent constraints which correspond to a Jacobian matrix of rank $M'$ on the constraint surface, and dependent constraints which hold as a consequence of the others. As a result of this, given the condition that a phase space function $G$ vanishes on the surface $\phi_{m}=0$ we can write it as 
\begin{equation}
G=g^{m}\phi_{m}
\end{equation}
for some functions of $g^{m}$. At the same time if for arbitrary variations $\delta q^{n}$, $\delta p_{n}$ we have
\begin{equation}
\lambda_{n} \delta q^{n}+\mu^{n}\delta p_{n}=0
\end{equation}
then 
\begin{equation}
\begin{array}{c}
\lambda_{n}=u^{m}\frac{\partial\phi_{m}}{\partial q^{n}}\\
\\
\mu^{n}=u^{m}\frac{\partial \phi_{m}}{\partial p_{n}}\\
\end{array}
\end{equation}
for some $u^{m}$. In the presence of redundant constraints the functions $u^{m}$ exist but are not unique. Now we can introduce the canonical Hamiltonian 
\begin{equation}
H=\dot{q}^{n}p_{n}-L
\end{equation}
The $\dot{q}$ only enter the Hamiltonian via the functions $p(q,\dot{q})$. This is a general property of the Legendre transformation. This Hamiltonian is not uniquely determined as a function of $p$ and $q$. This is clear because $\delta p_{n}$ are not all independent but are restricted to the primary constraints. The Hamiltonian is therefore defined only on the submanifold defined by the primary constraints but can be extended in arbitrary fashion outside that manifold rendering the theory unchanged to the transformations
\begin{equation}
H\rightarrow H+c^{m}(q,p)\phi_{m}
\end{equation}
Arbitrary variations in position and velocity induce a change in the Hamiltonian of the form 
\begin{equation}
\begin{array}{c}
\delta H=\dot{q}^{n}\delta p_{n}+\delta \dot{q}^{n}p_{n}-\delta\dot{q}^{n}\frac{\partial L}{\partial \dot{q}^{n}}-\delta q^{n}\frac{\partial L}{\partial q^{n}}=\\
\\
=\dot{q}^{n}\delta p_{n}-\delta q^{n}\frac{\partial L}{\partial q^{n}}
\end{array}
\end{equation}
where $\delta p_{n}$ can be seen as a linear combination of $\delta q$ and $\delta \dot{q}$. We can continue by rewriting 
\begin{equation}
(\frac{\partial H}{\partial q^{n}}+\frac{\partial L}{\partial q^{n}})\delta q^{n}+(\frac{\partial H}{\partial p_{n}}-\dot{q}^{n})\delta p_{n}=0
\end{equation}
and therefore 
\begin{equation}
\begin{array}{c}
\dot{q}^{n}=\frac{\partial H}{\partial p_{n}}+u^{m}\frac{\partial \phi_{m}}{\partial p_{n}}\\
\\
-\frac{\partial L}{\partial q^{n}}|_{\dot{q}}=\frac{\partial H}{\partial q^{n}}|_{p}+u^{m}\frac{\partial \phi_{m}}{\partial q^{n}}\\
\end{array}
\end{equation}
Using the first relation now we can obtain the velocities $\dot{q}^{n}$ from the knowledge of the momenta, and of extra parameters $u_{m}$. These extra parameters are providing us with a frame by being identified with coordinates on the surface of inverse images of a given momentum. 
The vectors $\frac{\partial \phi_{m}}{\partial p_{n}}$ are independent on $\phi_{m}=0$ due to the independence of the constraints. Therefore the relation between the set of $u$ and the velocities becomes one-to-one. Therefore we can write the parameters $u$ as functions of the coordinates and velocities by solving 
\begin{equation}
\dot{q}^{n}=\frac{\partial H}{\partial p_{n}}(q,p(q,\dot{q}))+u^{m}(q,\dot{q})\frac{\partial \phi_{m}}{\partial p_{n}}(q,p(q,\dot{q}))
\end{equation}
The Legendre transform from the $(q,\dot{q})$ space to the surface $\phi_{m}(q,p)=0$ of the $(q,p,u)$ space is defined by 
\begin{equation}
\begin{array}{c}
q^{n}=q^{n}\\
\\
p_{n}=\frac{\partial L}{\partial \dot{q}^{n}}(q,\dot{q})\\
\\
u^{m}=u^{m}(q,\dot{q})\\
\end{array}
\end{equation}
which brings us to 
\begin{equation}
\begin{array}{c}
q^{n}=q^{n}\\
\\
\dot{q}^{n}=\frac{\partial H}{\partial p_{n}}+u^{m}\frac{\partial \phi_{m}}{\partial p_{n}}\\
\\
\phi_{m}(q,p)=0\\
\end{array}
\end{equation}
which is now a map between spaces of the same dimension rendering the invertibility of the Legendre transformation even when $det(\frac{\partial^{2} L}{\partial \dot{q}^{n}\partial \dot{q}^{n'}})=0$ with the cost of adding extra variables. 
With this we can write the original Lagrangian equations in the equivalent Hamiltonian form
\begin{equation}
\begin{array}{c}
\dot{q}^{n}=\frac{\partial H}{\partial p_{n}}+u^{m}\frac{\partial \phi_{m}}{\partial p_{n}}\\
\\
\dot{p}_{n}=-\frac{\partial H}{\partial q^{n}}-u^{m}\frac{\partial \phi_{m}}{\partial q^{n}}\\
\\
\phi_{m}(q,p)=0
\end{array}
\end{equation}
Now, the Hamiltonian equations of motions can be derived from the variational principle 
\begin{equation}
\delta\int_{t_{1}}^{t_{2}}(\dot{q}^{n}p_{n}-H-u^{m}\phi_{m})=0
\end{equation}
for arbitrary variations in $q^{n}$, $p_{n}$ and $u_{m}$
The variables $u^{m}$ used to make the Legendre transformation invertible appear now as Lagrange multipliers for the primary constraints. 
The resulting equations of motion being 
\begin{equation}
\dot{F}=[F,H]+u^{m}[F,\phi_{m}]
\end{equation}
where 
\begin{equation}
[F,G]=\frac{\partial F}{\partial q^{i}}\frac{\partial G}{\partial p_{i}}-\frac{\partial F}{\partial p_{i}}\frac{\partial G}{\partial q^{i}}
\end{equation}

\section{what Madelung equations miss to become quantum?}
A simple stochastic approach to quantum mechanics must ultimately fail. This is so because quantum mechanics deals with probability amplitudes (or pre-probabilities) in which information about non-realised physical states is allowed to have a physical impact. Any stochastic approach would implicitly describe probabilities of outcomes that will ultimately be realised before they can contribute to physical reality, which makes usual stochastic approaches simply incompatible to quantum mechanics. 
Madelung equations appear to be very close to Schrodinger's equations and to a correct representation of quantum mechanics, however they ultimately fail in doing so due to some fundamental reasons. First, I have to mention that this is not a proposal on emergence of quantum dynamics from some classical substructure. I showed in the introduction that such an emergence is meaningless and I will dedicate a whole chapter for that subject later on. However, one aspect of the Madelung equation is that it does not provide a unique time evolution from initial data. It also requires a constraint that involves a discrete symmetry at every moment in time in order for it to provide physically acceptable solutions. To make it more explicit, we can write them as
\begin{equation}
\begin{array}{c}
m(\frac{\partial v}{\partial t}+(v\cdot \nabla)v)=-\nabla V+\frac{\hbar^{2}}{2m}\nabla\frac{\Delta\sqrt{\rho}}{\sqrt{\rho}}\\
\frac{\partial \rho}{\partial t}+\nabla\cdot (\rho v)=0\\
\end{array}
\end{equation}

which can be re-written as
\begin{equation}
\begin{array}{c}
\frac{\partial v}{\partial t}+v\cdot\nabla v=-\frac{1}{m}\nabla(Q+V)\\
\frac{\partial \rho}{\partial t}+\nabla\cdot(\rho v)=0\\
\end{array}
\end{equation}
where $v$ is the flow velocity, $\rho=m|\psi|^{2}$ is the mass density, $V$ is the potential of the problem as it appears in the Schrodinger's equation and
\begin{equation}
Q=-\frac{\hbar^{2}}{2m}\frac{\nabla^{2}\sqrt{\rho}}{\sqrt{\rho}}
\end{equation}
is the so called "Bohm potential". This potential has a worse reputation than what it actually means. It was assumed by Bohm to be part of his "guiding wave" theory, on which I do not insist because it ultimately was proven to be wrong. However, it also appears naturally from the Schrodinger equation 
\begin{equation}
i\hbar\frac{\partial \psi}{\partial t}=(-\frac{\hbar^{2}}{2m}\nabla^{2}+V)\psi
\end{equation}
To see how, we have to pick a solution of the Schrodinger equation in a polar form with a complex phase
\begin{equation}
\psi=A\cdot e^{\frac{i}{\hbar}S}
\end{equation}
This can therefore can be separated into an imaginary and a real part leading to the continuity equation and the quantum Hamilton Jacobi equation. The continuity equations appears from the imaginary part of the Schrodinger equation like
\begin{equation}
\frac{\partial A}{\partial t}=-\frac{1}{2m}[A\nabla^{2}S+2\nabla A\cdot \nabla S]
\end{equation}
and if we set $\rho=A^{2}$ we obtain the equation in the usual form 
\begin{equation}
\frac{\partial \rho}{\partial t}+\nabla\cdot(\rho v)=0
\end{equation}
given $\rho$ the probability density, and the velocity field $v=\frac{1}{m}\nabla S$. 
From the real part of the Schrodinger equation in polar form we obtain the Hamilton Jacobi equation expressed already for quantum systems as 
\begin{equation}
\frac{\partial S}{\partial t}=-[\frac{|\nabla S|^{2}}{2m}+V+Q]
\end{equation}
where the additional term, interpreted as a quantum modification of the Hamilton Jacobi equation is 
\begin{equation}
Q=-\frac{\hbar}{2m}\frac{\nabla^{2}A}{A}
\end{equation}
which seems to depend on the curvature of the amplitude of the wave function. 
At this point it is interesting to go back to the series of examples of when gauge variant observables are relevant. It is interesting to note that in general relativity, all gauge invariant observables must be non-local. This turns all observables of general relativity into non-localisable observables. This is the standard interpretation. Of course, in the case in which the curvature is small enough we can make approximations that render the gauge invariant observables approximately local, however, the non-trivial curvature doesn't permit to localise gauge invariant observables in absolute terms. A similar situation occurs in quantum chromodynamics only that there the "non-localisability" doesn't occur in spacetime, as with the case of general relativity. While in general relativity we have spacetime non-localisation of the observables, in QCD we have the same thing only the non-localisation occurs in the gauge space. This is why QCD and other non-gravitational gauge theories are fully localised in spacetime. However, the gauge space non-locality has an impact. In QCD the curvature is encoded by the $F_{\mu\nu}$ tensor which constructs the gauge field and the gauge connections and we have $F=dA+A\wedge A$. In general relativity, the curvature has a similar role. The only difference is that in the case of QCD the asymptotic freedom makes the curvature irrelevant at higher energies, leading to the approximate gauge invariance of "localised" (in the gauge space) colour. 
If we interpret quantum mechanics in a similar way as resulting from a gauge degree of freedom, we may expect to have a similar situation, namely for gauge "variant" observables, meaning, non-determined observables that would yield several possible values as outcomes, to become approximately gauge invariant when the "quantum curvature" becomes small. The term emerging in the $Q$ term above is determined by the two properties: the Planck constant, and the quantum curvature. It seems like not only the limit in which $\hbar\rightarrow 0$ renders a theory semi-classical, but also the case in which the amplitude of the curvature becomes small. This is very much in agreement with the observation that rendering the gauge connection almost flat makes the observables trivially gauge invariant (at least in an approximate sense). 
Now, to be sure, the "quantum" Hamilton Jacobi equation looks like a gauge extension of the original Hamilton Jacobi equation, but it is not truly quantum itself. We still need to postulate the role of the wavefunction as a probability amplitude, but this problem is solved if we go through the actual derivation of the Hamilton Jacobi equations and their solutions for a gauge system, because in that way we will obtain the indeterminacy required precisely by demanding the system to be gauge i.e. to contain arbitrary functions. 
The basic idea of the Hamilton Jacobi equation is to find those canonical transformations that would lead to a modified Hamiltonian that would vanish as it would only describe the evolution of constants of motion. The generating function of the transformation that would lead to that would be a complete integral. 
Basically, the equations of motion represent a canonical transformation between the coordinates and momenta at a given time $(q^{i},p_{i})(t)$ and the initial coordinates and momenta $(q_{0}^{i},p_{i}^{0})(t_{0})$. Instead of that, let us consider the transformation $(q^{i},p_{i})\rightarrow (\alpha^{i},\beta_{i})$ which is considered also to be canonical, and described by a generating function $S(q^{i}, \alpha^{i},t)$. With that we have
\begin{equation}
\begin{array}{cc}
p_{i}=\frac{\partial S}{\partial q^{i}}\;\\
\beta_{j}=-\frac{\partial S}{\partial \alpha^{j}}
\end{array}
\end{equation}
with the condition
\begin{equation}
det\frac{\partial ^{2}S}{\partial \alpha^{j}\partial q^{i}}\neq 0
\end{equation}
the variables $\alpha^{i}$ and $\beta_{i}$ being constants of motion. Hence the transformed Hamiltonian depending only on constants of motions, is basically vanishing as it only describes the evolution of constants of motion
\begin{equation}
\bar{H}(\alpha^{i},\beta_{i})=H+\frac{\partial S}{\partial t}
\end{equation}
which leads to the Hamliton Jacobi equation
\begin{equation}
\frac{\partial S}{\partial t}+H(q^{i},\frac{\partial S}{\partial q^{i}})=0
\end{equation}
The solution of this equation is the generating function $S(q^{i},\alpha^{i})$ depending on $n$ variables $\alpha^{i}$ and satisfying the determinant equation above is called a complete integral. A complete integral provides us with a general solution of the equations of motion
\begin{equation}
\beta_{j}=-\frac{\partial S}{\partial \alpha^{j}}
\end{equation}
allowing us to write $q^{i}=q^{i}(\alpha^{j},\beta_{k},t)$. 
If the time independent canonical variables $(\alpha^{i},\beta_{j})$ are the same as the initial conditions $(q_{0}^{i},p_{j}^{0})$ then the generating function is called the Hamilton principal function and is the same as the classical action 
\begin{equation}
W(q^{i},q_{0}^{i},t)=\int_{q_{0}^{i},t_{0}}^{q^{i},t} du L(q,\dot{q})
\end{equation}
where the integral is evaluated along the solution of the equation of motion between the two points. If the solution of the Hamilton Jacobi equation depends on fewer integration constants we have so called incomplete integrals. Those emerge by setting a part, say $m$ of the $\alpha$ to be equal to a definite value, say zero, and leave the other ones $\alpha^{A}$ variable. What happens is that because those variables have been set to zero, they will disappear from the integral $S$ and hence the dependence on them will be lost. Because of that the conjugate variables will be rendered undetermined 
\begin{equation}
\beta^{a}=-\frac{\partial S}{\partial \alpha_{a}}
\end{equation}
Therefore the remaining equations 
\begin{equation}
\begin{array}{c}
p_{i}=\frac{\partial S}{\partial q^{i}}\\
\beta_{A}=-\frac{\partial S}{\partial \alpha^{A}}\\
rank(\frac{\partial^{2}S}{\partial \alpha^{A} \partial q^{i}})=n-m\\
\end{array}
\end{equation}
will only provide an integral $S(q^{i},\alpha^{A},t)$ which can no longer determine a unique solution of the equations of motion. The remaining constants of motion $\alpha^{A}, \beta_{A}, \alpha_{a}=0$ do not characterise a single classical trajectory anymore. They do however characterise all trajectories that have the same values of the variables above but may have different values of the unknown conjugates $\beta^{\alpha}$. 
Starting on the solution of the equations of motion above, where the trajectories are not fully determined, at a point $(q^{i},p_{i})$ at a time $t$ and considering a variation that would bring us at a time $t+\delta t$ at forward in the phase space in a point that also satisfies those equations $(q^{i}+\delta q^{i}, p_{i}+\delta p_{i})$ then we would have
\begin{equation}
\begin{array}{c}
\delta q^{i}=\frac{\partial H}{\partial p_{i}}(q,p)\delta t\\
\delta p_{i}=-\frac{\partial H}{\partial q^{i}}(q,p)\delta t\\
\end{array}
\end{equation}
when $(q^{i},p_{i})$ and $(q^{i}+\delta q^{i}, p_{i}+\delta p_{i})$ give the same values for the unobserved conjugate momenta $\beta^{\alpha}$ and therefore lie on the same classical trajectory. This would be the case if we had solved the Hamilton-Jacobi theory for a complete integral but this must not be so here. The evolution in phase space satisfying the equations above for the incomplete integrals may produce different values for $\beta^{a}$. Since the shifts in $\beta^{a}$ are generated by the conjugate variables $\alpha_{a}$ in general one obtains
\begin{equation}
\begin{array}{c}
\delta q^{i}=[\frac{\partial H}{\partial p_{i}}(q,p)+\lambda^{a}\frac{\partial \alpha_{a}}{\partial p_{i}}(q,p)]\delta t\\
\delta p_{i}=[-\frac{\partial H}{\partial q^{i}}(q,p)-\lambda^{a}\frac{\partial \alpha_{a}}{\partial q_{i}}(q,p)]\delta t\\
\end{array}
\end{equation}
for some $\lambda^{a}$ with $\alpha_{a}(q,p)=0$. The parameter $\lambda$ establishes the evolution on the classical trajectory (if $\lambda=0$) or otherwise (if $\lambda \neq 0$). Quantum mechanics appears when we consider all integration constants disappearing from $S(q^{i},t)$. In that case we have reduced the determination of a path to a complete set of commuting conserved quantities with the possibility that they differ in all their conjugates realised. Interestingly enough, the Hamilton Jacobi theory of constraint systems is exactly the same as the one for incomplete solutions and the one for quantum mechanics. 
For constrained systems, if we describe the constraint surface as $G_{a}=0$ and we identify the variables we want to eliminate from the generating function $S$, namely $\alpha_{a}$ with an abelian representation of this constraint surface, then the conjugate variables $\beta^{a}$ are pure gauge. However, the same also represent the undetermined variables of quantum mechanics. There is basically no difference now between the gauge degrees of freedom and the quantum degrees of freedom. The distinction between them is purely conventional, and in fact, not even properly described even as a convention. As it is expected, any physical system has a series of compatible observables, forming a complete set of so called "gauge invariant functions", these "gauge invariant functions" are the equivalent to a complete set of compatible observables in quantum mechanics. This set is of course not unique, one can move around and replace observables in that set with other observables, with the condition of the set being complete and the observables to commute with each other. Those will be given by $(\alpha^{A},\beta_{A})$ commuting with $\alpha_{a}$. Those can be seen as canonical coordinates on the reduced phase space. The generating function $S(q^{i},\alpha^{A},\alpha_{a},t)$ defines the canonical transformation 
\begin{equation}
(q^{i},p_{i})\rightarrow \alpha^{A},\beta_{A},\alpha_{a},\beta^{a}
\end{equation}
such that the constraints will be $\alpha_{a}=0$. With those constraints satisfied one obtains a function $S(q^{i},\alpha^{A},t)$ that satisfies
\begin{equation}
\begin{array}{c}
G_{a}(q^{i},\frac{\partial S}{\partial q^{i}})=0\\
\frac{\partial S}{\partial t}+H_{0}(q^{i},\frac{\partial S}{\partial q^{i}})=0\\
rank(\frac{\partial^{2}S}{\partial \alpha^{A}\partial q^{i}})=n-m\\
\end{array}
\end{equation}
The information on $\frac{\partial S}{\partial \alpha_{a}}=-\beta^{a}$ is lost and the conjugate variable $\beta^{a}$ becomes arbitrary. We see here the emergence of the same structure we find in Feynman's path integrals. In particular the Feynman path integral is designed to integrate over all undetermined parameters defining a series of paths that are not physically realised but are in a sense physically distinct and possible to be realised. Gauge theory usually claims that the other paths are not physical, and are considered just a "redundancy" of the description, differing only in the values of undetectable parameters. However, the fact that those parameters are undetectable doesn't stop them from affecting the probability amplitudes and to generate probability distributions with unusual (from a classical point of view) shapes. The same thing is valid here, in the generation of several equivalent paths that however do play a role in the proper definition of gauge connections and of interactions. We obtained above the equations of the Hamilton Jacobi theory for constrained systems, but the constraint structure is the same that allows also for a Feynman path integral description of quantum mechanics. 
If we choose a point $(q^{i}(t),p_{i}(t))$ as a solution for
\begin{equation}
\begin{array}{c}
p_{i}=\frac{\partial S}{\partial q^{i}}\\
\beta_{A}=-\frac{\partial S}{\partial \alpha^{A}}\\
\end{array}
\end{equation}
we can obtain 
\begin{equation}
\begin{array}{c}
\dot{q}^{i}=[q^{i},H_{0}]+\lambda^{a}[q^{i},G_{a}]\\
\dot{p}^{i}=[p^{i},H_{0}]+\lambda^{a}[p^{i},G_{a}]\\
G_{a}(q,p)=0\\
\end{array}
\end{equation}
The generating function or action emerging from this formulation of the Hamilton-Jacobi theory contains all solutions of the problem that are gauge-related. In the same way we could say that it contains also all quantum trajectories that are to be integrated over, in which the choice of a frame amounts to a choice of unobservable, undetermined intermediary states. Therefore the overall structure given by this integral is "gauge invariant" per se, and making the theory consistent, both as a gauge and as a quantum theory. 
It is a matter of convention only, to call here the solutions of the above equations for $S(q^{i},\alpha^{A},t)$ that also satisfies the $rank=n-m$ equation a "complete solution" of the Hamilton Jacobi equation with constraints. This convention emerged from the assumption that the pure gauge part is physically irrelevant, describing only gauge related trajectories. But we can easily see that those are just the incomplete solutions of the previous formulation in which we re-evaluated the constraints. Is there a physical distinction between this an the situation in which we would "claim" (wrongly so) that the irrelevant, physically unrealised, intermediary states of a quantum system are "physically irrelevant"? Apparently not. In this case, the convention says that the incomplete solutions contain trajectories that are "physically distinguishable" in the sense that they differ in some gauge invariant quantities. The continuous elimination of those, leads again to a particular solution $S(q^{i},t)$ that depends on no integration constants at all and are the origin of quantum mechanics. The only difference is in what we "decide" to be physically distinguishable and what not. But this conventional distinction is in some sense meaningless. Here I have to refer again to the question of colour. If it is indeed gauge variant (which it is) then it should be also considered physically meaningless, but then I wonder how a chromodynamics would emerge? The same remains valid for simple choices of "frame variant" quantities even in special relativity, with length, time intervals, etc. which are all of course "frame dependant" but then not less physical. The fixed integration constants $\alpha_{a}$ are set equal to a fixed value, say zero, either by choice or because of some gauge invariance principle. They do not appear to have any distinct meaning, and makes us think that indeed we can re-arrange and choose what observables to be gauge invariant and what not, in the same way in which we can choose what set of observables is to be complete and commuting and what not. In quantum mechanics we do that all the time, we in fact may choose various complete commuting sets of observables to describe the same system. The determination of $\beta^{a}$ is also a matter of convention. There is no difference between the unknown $\beta^{a}$ or the "pure gauge" $\beta^{a}$ which makes them fundamentally the same. 
With linear homogeneous constraints in the momenta
\begin{equation}
G_{a}=a_{a}^{i}(q)p_{i}
\end{equation}
we obtain 
\begin{equation}
a_{a}^{i}(q)\frac{\partial S}{\partial q^{i}}=0
\end{equation}
which then doesn't mix $q$ and $p$ in the gauge transformation of one of them
\begin{equation}
q^{i}\rightarrow q^{i} + \epsilon^{a}a_{a}^{i}(q)
\end{equation}
and therefore it defines an "internal gauge symmetry". The gauge invariance of the action simply shows that $S(q+\delta_{\epsilon}q)-S(q)=0$ or $\delta_{\epsilon}S=0$. Of course, if we extend the constraint formulation to 
\begin{equation}
G_{a}=p_{a}-\frac{\partial V}{\partial q^{a}}(q)
\end{equation}
one finds that 
\begin{equation}
\delta_{\epsilon}S=S(q+\delta_{\epsilon} q)-S(q)-[V(q+\delta_{\epsilon} q)-V(q)]
\end{equation}
with $S$ transforming inhomogeneously. If the constraints are not linear one may not be able to construct a gauge transformation for $S$. In quantum mechanics however this non-linearity is represented in forms of linear operators acting on the wavefunction, in which $G_{a}$ appears as a generator and we transfer the gauge invariance on the wavefunction. The generator can now take various forms, even more general than the ones presented above, leading to the wavefunction gaining additional structure, beyond that of a scalar. This is of course very convenient, but it would not happen if gauge wouldn't have the same origin as the quantum phase. 
The Hamilton principal function $W(q_{2}^{i}, t_{2},q_{1}^{i},t_{1})$ can then be extended with the constraints (which appeared in an exactly similar manner as the gauge degrees of freedom, as we can remember from the previous discussions)
\begin{equation}
W(q_{2}^{i}, t_{2},q_{1}^{i},t_{1})=\int_{(q_{1},t_{1})}^{(q_{2},t_{2})}(p_{i}dq^{i}-H_{0} dt-\lambda^{a}G_{a}dt)
\end{equation}
The conventional approach identifies those functions as identical on gauge related paths, but this only happens because we assumed there would be no physical distinction between them, not because those values would really be identical. If that assumption is abandoned, we obtain basically the action that appears in the formulation of Feynman's quantisation prescription. The question is how strong the convention of "non-physicality" may be. Otherwise, as in quantum mechanics, in gauge theory, we have to ask ourselves what is "real" and if something is not real or really physically realised, can it have an impact on the theory or on other physical properties? We noticed that the way we introduced gauge invariance is exactly the same in which we introduced quantum intermediary states. We just claim that ones are "physically indistinguishable" while the others are "distinguishable", but mathematically they are both distinguishable. The addition of auxiliary variables in the gauge theory makes them clearly distinguishable and therefore we could imagine a larger space in which gauge related trajectories could play a physical role. It is therefore not a surprise that the procedure of quantisation interferes massively with the prescription of gauge invariance generating various anomalies. It is also not surprising that the resolution of the anomalies happens to be by adding supplemental variables or making higher-dimensional extensions. We can now easily obtain 
\begin{equation}
p_{i}(t_{2})=\frac{\partial W}{\partial q^{i}(t_{2})}
\end{equation}
and the expansion 
\begin{equation}
\frac{\partial W}{\partial t_{2}}=-(H_{0}+\lambda^{a}G_{a})(q_{2},\frac{\partial W}{\partial q_{2}})
\end{equation}
Once we decide to move on one trajectory, we can consider the other ones to be non-physical, which changes the problem into one that doesn't mix $q_{2}^{i}$ with $q_{1}^{i}$ but this is expected both in gauge theory (gauge invariance of physical solutions) and in quantum mechanics (the possible but physically non-realised solutions, once one is factually chosen, generates the "collapse" of the others). Of course, in quantum mechanics the collapse is seen as an update of the maximal knowledge about the system. However, before this happens the constraint has to be imposed, and results in a modified form of the Hamilton Jacobi equation. It is interesting to note that a similar correction emerges also in a quantum approach to Hamilton Jacobi, when one starts with a description that contains a wavefunction and one arrives at the so called "quantum potential" which, although seems to be unrelated to the gauge dependence constraint, results in a term that has precisely the same effect on the possible trajectories 
\begin{equation}
\begin{array}{c}
\frac{\partial W}{\partial t}=-[\frac{|\nabla S|^{2}}{2m}+V+Q]=-[H_{0}+Q]\\
\\
Q=-\frac{\hbar^{2}}{2m}\frac{\nabla^{2}A}{A}\\
\end{array}
\end{equation}
Now, one has to remark some things: first, in the quantum Hamilton Jacobi equation this "potential" is obtained from the real part of the Schrodinger equation after performing a polar decomposition of the wavefunction and is not apparently related to the Hamiltonian itself. It also involves non-local properties and seems to be derived from the curvature of the amplitude of the wavefunction. However, this construction is plagued by the guiding wave interpretation of Bohm. That interpretation is of course not necessary if one notices that the emergence of such a potential as a gauge contribution is what actually happens. Indeed, in a gauge approach, such a potential would not emerge as a property of the dynamical system, other than that of allowing for a gauge, and would therefore allow for global effects as those we see in quantum mechanics, however, without the problems of seeing actually realised independent trajectories and facing various problems like the "many world" interpretation. 
In the gauge interpretation we do not yet have a wavefunction and hence we do not have the amplitude of a wavefunction $A$. However, we can regard the gauge structure as an emerging fibre bundle, in which the interaction or the potential is seen as a connection on the fibres with a curvature. The gauge space curvature involves already global information but usually one obtains the connection as a covariant derivative that restores gauge invariance disturbed by the presence of a curvature form over the fibres. It appears that quantum mechanics itself appears as a form of curvature that gives access to global information on the gauge bundle and accesses global information in the same way in which all other interactions do. However, it accesses the curvature in a space of non-realised intermediate states, which allows for correlations that would not exist in classical physics.

First we notice that the introduction of a gauge invariance is not different from the introduction of a quantum modification. Second, in terms of interpretation we notice that the quantum expression provides us with the same type of non-measurable choice given by an arbitrary function which propagates alongside the solution as the one we recover in the Madelung equation. The only difference between the two is that this type of arbitrariness is encoded in the Madelung equation at the level of the equation, while in usual quantum mechanics, this is part of the complex phase which is encoded in the Ansatz that made us define the wavefunction as we did. There is essentially no difference up to now. There is however another constraint that is required for the Madelung equation to be defined as a description of quantum mechanics. In particular a rotational symmetry that has to be implemented at each point in time. This is nothing but the requirement of a gauge invariance making those solutions truly quantum. It is important to note that in general a usual symmetry does not play the role of a gauge symmetry. A physical symmetry is related to the formulation of the theory or to some object in it. We may have some discrete symmetry in the lattice of crystallin solid, etc. That would not be a gauge symmetry. A gauge symmetry needs to be introduced like an arbitrary function which we can choose in whatever way we wish at any moment in time. This freedom introduces a set of constraints on the variables of the theory, making them not all independent. The same thing occurs in quantum mechanics with the introduction of the phase, with the only difference that the Madelung equation appears via a separation of the two parts, therefore the gauge symmetry is split between the two, and must be restored, which is what the additional constraint does. Therefore it is interesting to note that the addition of a gauge symmetry and the required constraints in the case of Madelung equation is equivalent to rendering the theory "truly quantum" i.e. in agreement with Schrodinger's equation. This further endorses the claim that gauge is actually quantum. 

The idea that quantum is gauge brings us to a series of new observations regarding the nature of gauge dualities and how they can be linked to quantum information. In fact, all elements of quantum information should have a dual in gauge theory, leading to a systematic approach to gauge dualities. 

\section{quantum as a curvature, but which?}
The history of physics reminds us of the evolution of our understanding of interactions in the following way. General relativity was constructed by noticing that spacetime has a curvature and that the relation between matter and geometry is a double arrowed map. Matter determines the curvature of spacetime and spacetime determines the dynamics of matter. The result of the introduction of curvature, which implies some form of global information (as we well know, a local observer cannot determine whether he is in a state of accelerate motion or under the effects of a gravitational field) results in the fact that "physical" gauge invariant observables of general relativity must be non-local. Any local gauge (diffeomorphism) invariant gravitational observable must be only approximate and appears only in the limit of small curvature. In our current understanding, the gravitational field appears as a correction in the connection (covariant derivative) on the fibre bundle that has as basis space spacetime. 
The non-locality of observables means that in order to have a non-ambiguous description of gravitational results we need access to information that is not localised. This is a direct result of the existence of curvature in spacetime. 
A similar situation occurs in QCD where the colour charge also is not gauge invariant. As this happens to correspond to a curvature in gauge space, and not directly in spacetime, QCD is a fully local theory, as are all the other gauge field theories. The non-local nature of QCD appears due to the rotations in gauge space. We cannot unambiguously define the colour charge, as we can arbitrarily choose different gauge transformations that will lead to different colours or to superpositions of those. However, the cause for this non-locality is also the existence of curvature of the associated gauge bundle fibres. Again, the theory becomes non-ambiguous only in the limit of small curvature in the gauge space. 
In quantum mechanics we are careful to specify that non-locality is not truly related to a causally triggered non-locality. In fact, this non-locality appears only as a result of statistics. Every measurement we perform in quantum mechanics is strictly local, however, the non-locality emerges in a strange way when we repeat certain experiments several times. Then we observe correlations that suggest that non-realised intermediate states do play a role, we also observe therefore statistical interferences by means of the shape of the obtained distribution functions, and basically we gain access to non-local information via a global structure, say, the rule of spin conservation in the example above. 
What we lack here however is a concept of curvature. The limit in which quantum effects are small are usually related to an approximation in which the Planck constant can be considered as small. A more general limit is one in which the dual curvature would be considered small, however, there could exist mixed situations in which the smallness of the Planck constant is compensated by the high dual "curvature" leading to quantum effects that are not strictly determined by the Planck scale. I am totally not surprised by the fact that we observe large scale quantum effects in a series of experiments nowadays. 
The problem is how can we identify what this curvature dual would be in the context of quantum mechanics? We know that it is only through a statistical interpretation of quantum mechanics that we can access the relative quantum phase of the components of a general wavefunction. Therefore the first idea would be that we need to identify some curvature in a space associated to statistics. This however leads to the same mistakes that have been made in the past. Finally it is relatively clear nowadays that we cannot replace quantum mechanics with a stochastic theory, no matter where (in the region of classically realised states) we place the stochasticity. The stochasticity or randomness of quantum mechanics is strictly a phenomenon linked to the statistics of measurements, and never to the actual state of the system, which is propagated by means of the Schrodinger equation. Each outcome by itself is well determined. We cannot possibly imagine some "brownian force" acting on a particle, changing its position or momentum arbitrarily. Also, we do not have the possibility of seeing a dynamical mechanism that leads to a specific outcome of a quantum experiment. This simply doesn't exist because the process is not a dynamical one, it is rather an informational one. The standard explanation is that the wavefunction as a catalogue of knowledge is updated by the new information that can be extracted by means of the extra degrees of freedom added by the measurement setup. In any case, this is not a question of dynamics as suggested by quite some research across the history. 
The fact that some form of curvature is involved in the appearance of quantum effects has been discussed previously, unfortunately in the context of a stochastic approach to quantum mechanics and the pilot wave formalism. In this article I do not wish to show the emergence of quantum mechanics from any classical degrees of freedom, nor to provide a classical to quantum duality. Basically, I do not even believe such a duality exists. Quantum mechanics describes a series of phenomena that are inaccessible to any classical approach, including those based on stochastic processes. However, the goal of this article is to establish a duality between gauge phenomena and quantum mechanics, and there, the concept of a curvature of a connection is relevant. 
In order to see what space is relevant as a base space on which we can construct a fibre with curvature, let us consider the dynamical problem. We will require to have the coordinates $q^{i}$ with the parallel transformations generated by $q^{i}\rightarrow q^{i}+dq^{i}$ and for the space in which we generate this connection we define a length 
\begin{equation}
ds^{2}=g_{ik}(q)dq^{i}dq^{k}
\end{equation}
However, in this case we have to work with trajectories over undetermined variables and their conjugates. In this case we are interested in a generally covariant formulation in which the evolution of the system is described by putting the canonical variables and time on equal footing. In practical cases however time is considered as a variable on which the other canonical variables depend, hence we automatically assumed that time has a physical significance (it is measurable) while the others may or may not be so. This viewpoint has been of course insufficient for a relativistic description and therefore has been largely abandoned. If we introduce time as a canonical variable together with the rest, and describe the correlations between the original dynamical variables and time we have to introduce an additional arbitrary parameter and therefore to enlarge the set of canonical variables. The new arbitrary parameter however doesn't have an independent physical significance and therefore we obtain a formalism that is invariant with respect to its reparametrisations. This means the formalism is generally covariant. Practically, generally covariant systems appear either by having an original system in which time was not a canonical variable, but which we parametrised in order to obtain a general covariance, or we had a system that was from start generally covariant (as is the case with the gravitational field in general relativity). It is in general preferable to have as generally covariant a system as possible because in general covariant theories the motion is just the "unfolding of a gauge transformation". 
Let us start with a canonical set of variables $(q^{i},p_{i})$, a hamiltonian $H_{0}(q,p)$, and no constraint. The action is
\begin{equation}
S[q^{i}(t),p_{i}(t)]=\int_{t_{1}}^{t_{2}}(p_{i}\frac{dq^{i}}{dt}-H_{0})dt
\end{equation}
If we set the time to be the zero position $t=q^{0}$ and the conjugate being $p_{0}$ and consider those to be canonical variables we have
\begin{widetext}
\begin{equation}
S[q^{0}(\tau),q^{i}(\tau), p_{0}(\tau), p_{i}(\tau),u^{0}(\tau)]=\int_{\tau_{1}}^{\tau_{2}}[p_{0}\dot{q}^{0}+p_{i}\dot{q}^{i}-u^{0}(p_{0}+H_{0})]d\tau
\end{equation}
\end{widetext}
The extremisation of both actions leads to the same classical motions.  If we extremise first the second action with respect to $u^{0}$ and $p_{0}$ producing $\gamma_{0}=p_{0}+H_{0}=0$ and $\dot{t}-u^{0}=0$. We obtain the reduced action depending only on $q^{\mu}(\tau)$ and $p_{i}(\tau)$
\begin{equation}
\int_{\tau_{1}}^{\tau_{2}}(p_{i}\dot{q}^{i}-H_{0}\dot{t})d\tau=\int_{t_{1}}^{t_{2}}(p_{i}\frac{dq^{i}}{dt}-H_{0})dt
\end{equation}
We remember that in generally covariant systems used to construct the path integrals we admit even solutions that go backwards in time even in the non-relativistic limit. The general covariant expression of the action above depends on an extra pair of canonical variables and is constrained by $\gamma_{0}=0$. The extended Hamiltonian therefore contains only the constraint. Therefore the motion is represented just by the evolution of the gauge transformation. Given additional gauge generators $\gamma_{a'}$, $(a'=1,...,m)$ and other second class constraints $\chi_{\alpha}=0$, we obtain the action
\begin{equation}
S=\int_{\tau_{1}}^{\tau_{2}}(p_{\mu}\dot{q}^{\mu}-H_{E})d\tau
\end{equation}
with 
\begin{equation}
H_{E}=u^{a}\gamma_{a}+u^{\alpha}\chi_{\alpha}
\end{equation}
with $a=0,...,m$. 
The extended action is invariant to the transformations
\begin{equation}
\begin{array}{c}
\delta q=\dot{q}\epsilon\\
\delta p=\dot{p}\epsilon\\
\delta u^{a}=\dot{(u^{a}\epsilon)}\\
\delta u^{\alpha}=\dot{(u^{\alpha}\epsilon)}\\
\end{array}
\end{equation}
with $\epsilon(\tau_{1})=\epsilon(\tau_{2})=0$. This reparametrisation differs from the usual gauge symmetries generated by $\gamma_{a}$ and is not an independent symmetry. Therefore the extended action with zero Hamiltonian is invariant under arbitrary reparametrization of the $\tau$ variable. In general if $q$ and $p$ transform like scalars under time reparametrisations the Hamiltonian is weakly zero for a generally covariant system. However a time dependent canonical transformation changes the value of the Hamiltonian. A situation would be to perform a canonical transformation that depends on $\tau$ and to obtain a non-zero Hamiltonian keeping the explicit reparametrisation invariance of the action. The new canonical variables will not be scalars, and the parameter $\tau$ itself is not transforming like a scalar. The canonical variables may transform inhomogeneously under reparametrisation as a connection. This analysis shows that there is yet another ambiguity in the description of a dynamical system. An arbitrary function of time in the general solution of the equations of motion makes it impossible to say if that arbitrary function appears because some canonical variable is not observable or because time itself is not observable, by using the equations of motion. When we look at Maxwell's equations we only assume that time is more physical and hence one component of the vector potential is not observable. In any case, what is interesting is that in a quantum formulation we can continue to reduce the integrals of motion, obtaining a series of undetermined paths which we have to consider as intermediate unobservable states of the system. If that is the case, a general covariant construction over those paths must involve transformations of the canonical coordinates by means of connections. With this happening in the extended gauge space we can now define a parallel transport in this inner space as in $A^{i}\rightarrow A^{i}+\delta A^{i}$ after a transformation $q^{i}\rightarrow q^{i}+\delta q^{i}$. We can define the transformation as
\begin{equation}
\delta A^{i}=\Gamma_{kl}^{i}A^{k}dq^{l}
\end{equation}
with the connections $\Gamma$ defining the type of curvature of the underlying space constructed by the unobservable variables. We demand that the parallel transport of the length of a vector 
\begin{equation}
l=(g_{ik}A^{i}A^{k})^{1/2}
\end{equation}
changes as a linear law
\begin{equation}
\delta l=l\phi_{k}dq^{k}
\end{equation}
with $\phi_{k}$ the covariant components of a vector in our gauge space. The resulting geometry is the Weyl geometry. Treating the connection as a gauge field we obtain a modification that may be expected to be of the form 
\begin{equation}
H\rightarrow H+\frac{\hbar^{2}}{2m}\cdot R(q,t)
\end{equation}
where $R(q,t)$ plays the role of a curvature as determined in ref [9]. However this appears strictly as resulting from a curvature in a gauge space that is constructed by means of Weyl type connections. It is important to notice that, as discussed in the approach to the duality between gauge and quantum, the respective trajectories are non-realised and non-determined, hence the standard interpretation of quantum mechanics is maintained. However, the presence of this curvature term in the gauge space produces a non-trivial effect on those unrealised and undetermined trajectories, effect that is created by the curvature of the trajectories themselves. This curvature is finally determined by the type of experimental setup and the specific choice of a set of compatible observables, as well as the non-realised observables (gauge variant ones) in the system. In ref [9] those are assumed to be happening in spacetime. I strongly suspect this is wrong. We are not, after all, talking about realised intermediate states in spacetime, hence we cannot have any effect in the actual spacetime, which is also why our theory would be strictly local in terms of spacetime, however, it would present a series of non-local properties once the statistics of the measurements is involved. The same curvature obtained in ref [9] should be valid, 
\begin{equation}
R=\dot{R}+(n-1)[(n-2)\phi_{i}\phi^{i}-2(\frac{1}{\sqrt{g}})\partial_{i}(\sqrt{g}\phi^{i})]
\end{equation}
where $g=det(g_{ik})$ and $\dot{R}$ is the Riemannian curvature of the gauge space. 
This underlines even more the fact that the Quantum Hamilton Jacobi equation makes sense only in the case of a gauge system in which we have an arbitrary function propagating along the solutions. There is no way in which the quantum correlations can be represented in normal spacetime in a local and causal manner, however, no such complications appear in gauge space particularly in a fully reparametrisation covariant construction. Trajectories moving backwards in time would be surprising if they occurred in spacetime, but as gauge trajectories in the gauge space of a fully reparametrisation covariant formulation they represent nothing special.
\section{conclusion}
I showed in this article that there are good arguments in favour of the assumption that Gauge is Quantum. With such a duality we definitely do not claim that quantum is classical or that there would ever be a one to one correspondence between classical and quantum physics. That would be in contradiction with various theoretical and experimental results. However, there are various applications of this result that would appear as extremely important. First, our causal structure is defined by means of gauge connections. The photon field is a result of a gauge connection and its speed limit defines the causal structure. An explanation for that type of causal constraint can be found by looking on the other side, in the quantum realm. In the same sense, we may expect to find duals between spontaneous gauge symmetry breaking as it is seen in the Higgs mechanism and some quantum information results. Moreover, new dualities may emerge between gauge field theories on one side and strongly entangled theories on the other. The spacetime causal structure itself may turn out to be dependent on various quantum characteristics making the ER-EPR duality even more strict.
Moreover, given that quantum effects depend not only on the Planck constant $\hbar$ but also on the curvature of the special connection in the gauge space, we may expect to see quantum effects not only in situations in which $\hbar$ is comparable with the scale of our system, but also in situations in which $\hbar$ is relatively small, but its small scale is compensated by a very large inner space curvature. Therefore we expect quantum effects to occur at large scales for example, or in systems in which such effects would be prohibited by the usual temperature/energy considerations. 
Such future research could have an impact on various areas of study, including the study of structured light, a subject of major interest particularly when considering non-linear effects and higher harmonic structured light [11], [12]. 
\section{Data Availability Statement}
Data sharing not applicable to this article as no datasets were generated or analysed during the current study.

\end{document}